\documentstyle[12pt]{article}
\input epsf

\begin{document}
\begin{titlepage}
\thispagestyle{empty}
\title{The $Q^2$ evolution of Soffer inequality}
\vspace{4.0cm}
\author{C. Bourrely, J. Soffer \\
\small \em Centre de Physique Th\'eorique - CNRS - Luminy,\\
\small \em Case 907 F-13288 Marseille Cedex 9 - France \\
  and  \\ O.V. Teryaev \\
\small \em Bogoliubov Laboratory of Theoretical Physics, \\
\small \em Joint Institute for Nuclear Research,\\
\small \em 141980 Dubna, Moscow region,  Russia}
\date{}

\maketitle

\begin{abstract}
DGLAP evolution equations may be presented in a form completely analogous
to the Boltzmann equation. This provides a natural proof of the positivity
of the spin-dependent parton distributions,
provided the initial distributions at $Q^2_0$ are
also positive. In addition, the evolution to $Q^2 < Q^2_0$
may violate  positivity, providing therefore a 'time arrow'.
The positivity condition is just $|\Delta P_{ij} (z)| \leq
P_{ij} (z) $ for $z < 1$ for all types of partons, while the
$'+'$prescription and terms containing $\delta(1-z)$ do not
affect positivity.
This method allows one to complete immediately the existing
proof of Soffer inequality at leading and next-to-leading order.
\end{abstract}
\vspace{1 cm}


\smallskip

\noindent Number of figures : 3

\smallskip

\noindent September 1997

\noindent CPT-97/P.3538

\noindent Web address: www.cpt.univ-mrs.fr
\end{titlepage}

The positivity constraints play an essential role in the
current analysis of various parton distributions, describing the
nucleon spin structure. One should first mention here the Soffer
inequality, putting an additional constraint for the
quark transversity distribution \cite{S}
\begin{eqnarray}
|h_1(x)| \leq q_+(x) \equiv {1 \over 2} [q(x) + \Delta q(x)].
\label{S}\end{eqnarray}

Also, the current next-to-leading (NLO)
parametrizations \cite{GS,GRSV} are chosen in such a way, that  positivity
for all helicity parton distributions is respected, i.e.
$|\Delta f (z, Q^2)| \leq f (z, Q^2)$.

One may wonder, to what extent the $Q^2$ evolution is compatible with
the positivity constraints. In the present paper it is shown,
that the relation is especially clear when the interpretation
of the evolution equation as a kinetic equation is
adopted. As an immediate consequence, the complete proof of
Soffer inequality at leading order (LO) and next-to-leading order
(NLO), is easily obtained.

The probabilistic understanding of the evolution equations
came already from the pioneering papers \cite{GL,AP,D}.
Actually, its standard form \footnote{For brevity, the
argument $t$ will not be written down explicitly in the parton densities.}
(for the time being, we shall confine ourselves to the non-singlet
(NS) case)
\begin{eqnarray}
{dq(x) \over{dt}}=
{\alpha_s \over {2 \pi}} \int_x^1 dy {{q(y)} \over y} P ({x \over y}),
\label{GLAP}\end{eqnarray}
may be interpreted as a 'time' $t=ln Q^2$  evolution  of the
'particles' density $q$ in the one dimensional space
$0 \leq x \leq 1$ due to the flow from the right to the left,
with the probability equal to the splitting kernel $P$.
The key element of such an interpretation is the problem of the
infrared (IR) singular terms in $P$, which was considered in detail
some years ago \cite{CQ} (see also \cite{DP}).
The kinetic interpretation is preserved
provided the $'+'$ form of the kernel is presented in the following way
\begin{eqnarray}
P_+ (z) =P(z)-\delta(1-z)\int_0^1 P(y) dy,
\label{+}\end{eqnarray}
leading to the corresponding expression for the evolution equation
\begin{eqnarray}
{dq(x) \over{dt}}=
{\alpha_s \over {2 \pi}}[ \int_x^1 dy {{q(y)} \over y}
P ({x \over y})- q(x) \int_0^1 P(z) dz].
\label{+e}\end{eqnarray}
This form has already a kinetic interpretation because
the second term in the square brackets describes the flow of
the partons at the point $x$ \cite{CQ}. It seems however
instructive \cite{OT} to make this resemblance even clearer
by the simple change of variables  $z=y/x$ in the second term, which
gives the equation the following symmetric form
\begin{eqnarray}
{dq(x) \over{dt}}=
{\alpha_s \over {2 \pi}}[ \int_x^1 dy {{q(y)} \over y}
P ({x \over y})- \int_0^x dy {q(x) \over x} P({y \over x}) ].
\label{b}\end{eqnarray}
It allows us to write it down in a way absolutely similar
to the Boltzmann equation, namely
\begin{eqnarray}
{dq(x) \over{dt}}=
\int_0^1 dy ( \sigma(y \to x) q(y)-\sigma(x \to y) q(x)),
\label{B}\end{eqnarray}
where the one-dimensional analogue of the Boltzmann
'scattering probability' is defined as
\begin{eqnarray}
 \sigma(y \to x) = {\alpha_s \over {2 \pi}} P ({x \over y})
 {{\theta (y > x)} \over y}.
\label{sigma}\end{eqnarray}

As a result, the known properties of the evolution equation
receive the following simple interpretation: the cancellation of
the IR divergencies between the contributions of the real and virtual
gluons emission is coming from the equality of in- and out- flows
when in both terms of (\ref{B}) one has $y \sim x$, following from
the continuity condition for the 'particles' number.

Also, the conservation of the vector current
\begin{eqnarray}
\int_0^1 dx {dq(x) \over{dt}}=
\int_0^1 dx dy (\sigma(y \to x) q(y)-\sigma(x \to y) q(x))=0,
\label{c}\end{eqnarray}
comes from the integration of an antisymmetric function in a symmetric
region.

Morever, one can easily question \cite{OT} the validity of the famous
Boltzmann H-theorem. It happens not to be valid because
of the violation of one of the key Boltzmann assumptions,
namely, 'molecular chaos' $\sigma(x \to y) = \sigma(y \to x)$.
The violation is due, in particular, to the $\theta-$function,
reflecting the causality properties, as it is well known.
 We shall also address this subject
when discussing positivity, to which we are now ready to come.

As the 'particles' density
in the Boltzmann equation is positive by definition,
this property is absolutely obvious.
The negative second term in (\ref{B}) cannot change the sign of the
distribution
because it is 'diagonal' in $x$, which means that it is proportional
to the function at the same point $x$, as in the l.h.s..
When the distribution gets too close to zero,
its stops decreasing. This is true for both $'+'$ and $\delta(1-z)$
terms, for any value of their coefficient (if it is positive,
it will reinforce the positivity of the distribution).

Let us consider now the spin-dependent case. For simplicity, we
postpone the discussion of quark-gluon mixing for a moment, but allow
the spin-dependent and spin-independent kernels to be different,
as they are at NLO.
It is most convenient to write down the equations for definite parton
helicities, which was actually the starting point in deriving
the equations for the spin-dependent quantities \cite{AP}.
Although the form, which we shall use, mixes the contributions of
different helicities, it makes the positivity properties especially
clear. So we have
\begin{eqnarray}
{dq_+(x) \over{dt}}=
{\alpha_s \over {2 \pi}} (P_{++} ({x \over y}) \otimes q_+(y)+
P_{+-} ({x \over y}) \otimes q_-(y)), \nonumber \\
{dq_-(x) \over{dt}}=
{\alpha_s \over {2 \pi}} (P_{+-} ({x \over y}) \otimes q_+(y)+
P_{++} ({x \over y}) \otimes q_-(y)).
\label{h}\end{eqnarray}

Here $P_{++}(z)=(P(z)+\Delta P(z))/2, P_{+-}(z)=(P(z)-\Delta P(z))/2$
are the evolution kernels for definite helicities, and the shorthand
notation for the convolution is adopted.
As soon as $x < y$, the positivity of the initial distributions
($q_+(x, Q_0^2), q_-(x, Q_0^2) \geq 0$, or
$|\Delta q (x, Q_0^2)| \leq q (x,Q_0^2)$)
is preserved, if both kernels
$P_{++},P_{+-}$ are positive, which is true, if
\begin{eqnarray}
|\Delta P (z)| \leq P (z), \ z < 1.
\label{ineq}\end{eqnarray}

The singular terms at $z=1$  are not altering positivity,
because they appear only in the diagonal (now in helicities)
kernel $P_{++}$ (only forward scattering is
IR dangerous). From the kinetic interpretation again
the distributions $q_+,~q_-$ stop decreasing, as soon as they
are close to changing sign.

Now to extend the proof to the quark gluon mixing is trivial.
One should write down the expressions for the evolutions of quark
and gluon distributions of each helicity
\begin{eqnarray}
{dq_+(x) \over{dt}}=
{\alpha_s \over {2 \pi}} (P_{++}^{qq} ({x \over y}) \otimes q_+(y)+
P_{+-}^{qq} ({x \over y}) \otimes q_-(y))  \nonumber \\
+P_{++}^{qG} ({x \over y}) \otimes G_+(y)+
P_{+-}^{qG} ({x \over y}) \otimes G_-(y),
\nonumber \\
{dq_-(x) \over{dt}}=
{\alpha_s \over {2 \pi}} (P_{+-} ({x \over y}) \otimes q_+(y)+
P_{++} ({x \over y}) \otimes q_-(y)) \nonumber \\
+P_{+-}^{qG} ({x \over y}) \otimes G_+(y)+
P_{++}^{qG} ({x \over y}) \otimes G_-(y),  \nonumber \\
{dG_+(x) \over{dt}}=
{\alpha_s \over {2 \pi}} (P_{++}^{Gq} ({x \over y}) \otimes q_+(y)+
P_{+-}^{Gq} ({x \over y}) \otimes q_-(y) \nonumber \\
+P_{++}^{GG} ({x \over y}) \otimes G_+(y)+
P_{+-}^{GG} ({x \over y}) \otimes G_-(y)),  \nonumber \\
{dG_-(x) \over{dt}}=
{\alpha_s \over {2 \pi}} (P_{+-}^{Gq} ({x \over y}) \otimes q_+(y)+
P_{++}^{Gq} ({x \over y}) \otimes q_-(y) \nonumber \\
+P_{+-}^{GG} ({x \over y}) \otimes G_+(y)+
P_{++}^{GG} ({x \over y}) \otimes G_-(y)).
\label{hs}\end{eqnarray}
If the inequality (\ref{ineq}) is valid for each type of partons 
\cite{BLT},
\begin{eqnarray}
|\Delta P_{ij} (z)| \leq P_{ij} (z), \ z < 1; \ i,j = q,G,
\label{ineqs}\end{eqnarray}
all the kernels, appearing in the r.h.s. of such a system, are
positive. Concerning the singular terms, they are again diagonal,
now in parton type, and do not affect positivity.
The validity of these equations at LO comes just from the
way they were derived, as the (positive) helicity-dependent
kernels were in fact first calculated in ref.\cite{AP}.
At  NLO, the situation is more controversial \cite{BLT}.

To conclude, the stability of positivity under $Q^2-$ evolution
comes from two sources:

i) the inequalities (\ref{ineqs}), leading to
the increasing of distributions,

ii) the kinetic interpretation
of the decreasing terms.

For the latter it is crucially important, that they
are diagonal in $x$, in helicity and in parton type, which is
related to their IR nature.

It is interesting to note \cite{OT}, that the 'back' evolution to
$Q^2 < Q^2_0$ (which is described by DGLAP equations as well,
 as soon as $Q^2 >> \Lambda_{QCD}^2$),
generally speaking, does not preserve positivity, as the regular and
singular terms change their role. Namely, all the terms for $ z < 1 $,
lead to the falloff of the distributions, while the singular terms
make them increase. Starting from a 'pathological'
positive distribution (say, increasing with $x$),
it is possible to get the violation of positivity.
This just means, that an arbitrary distribution cannot be obtained as
a result of the evolution starting from lower $Q^2$ .
As a result, the evolution kernels in fact impose some restrictions
on the boundary conditions (mainly coming from the mentioned
causality properties,
defining some 'time arrow'),
which are not easy to see in the original renormalization group
equations for the moments.

One should note in this connection,
that the inequality analogous to (\ref{ineqs}) for the moments
of the splitting kernels (anomalous dimensions)
is not sufficient for positivity, since the
distribution with all the positive moments may be negative for small $x$.

Let us now come to the evolution of Soffer inequality. According to
the previous analysis it is straightforward to define the following
'super'-distributions
\begin{eqnarray}
Q_+(x) =   q_+(x) +  h_1(x), \nonumber \\
Q_-(x) =   q_+(x) -  h_1(x).
\label{Q}\end{eqnarray}
Due to Soffer inequality, both these distributions are positive
at some point $Q^2_0$, and the evolution equations for the NS case
take the form
\begin{eqnarray}
{dQ_+(x) \over{dt}}=
{\alpha_s \over {2 \pi}} (P^Q_{++} ({x \over y}) \otimes Q_+(y)+
P^Q_{+-} ({x \over y}) \otimes Q_-(y)), \nonumber \\
{dQ_-(x) \over{dt}}=
{\alpha_s \over {2 \pi}} (P^Q_{+-} ({x \over y}) \otimes Q_+(y)+
P^Q_{++} ({x \over y}) \otimes Q_-(y)), \nonumber \\
\label{eQ}\end{eqnarray}
where the 'super'-kernels at LO are just
\begin{eqnarray}
P^Q_{++}(z) & \equiv & {[P_{qq}^{(0)}(z) + P_h^{(0)}(z)] \over 2}
\nonumber \\
& = &{C_F \over 2} [{{(1+z)^2} \over{(1-z)_+}} + 3\delta (1-z)],
\nonumber \\
P^Q_{+-}(z) &\equiv &{[P_{qq}^{(0)}(z) - P_h^{(0)}(z)] \over 2} \\
&=&{C_F \over 2} (1-z) \nonumber.
\label{PQ}\end{eqnarray}

One can easily see, that the inequalities analogous to (\ref{ineqs})
are satisfied, so that both $P^Q_{++}(z)$ and $P^Q_{+-}(z)$
are positive for $z<1$, while the singular term does appear
only in the diagonal kernel.
So, both requirements are valid and Soffer inequality is preserved
under LO evolution. The extension to the singlet case is trivial,
as the chiral-odd transversity distributions does not mix with gluons.
Therefore, they affect only the evolution of quarks, and lead to the
presence in the r.h.s. of the same extra terms as in (\ref{hs})
\begin{eqnarray}
{dQ_+(x) \over{dt}}=
{\alpha_s \over {2 \pi}} (P^Q_{++} ({x \over y}) \otimes Q_+(y)+
P^Q_{+-} ({x \over y}) \otimes Q_-(y) \nonumber \\
+P_{+-}^{qG} ({x \over y}) \otimes G_+(y)+
P_{++}^{qG} ({x \over y}) \otimes G_-(y)),
\nonumber \\
{dQ_-(x) \over{dt}}=
{\alpha_s \over {2 \pi}} (P^Q_{+-} ({x \over y}) \otimes Q_+(y)+
P^Q_{++} ({x \over y}) \otimes Q_-(y) \nonumber \\
+P_{+-}^{qG} ({x \over y}) \otimes G_+(y)+
P_{++}^{qG} ({x \over y}) \otimes G_-(y)),
\label{PQ1}\end{eqnarray}
which are all positive and free from singular terms, so positivity is
preserved.

\begin{figure}
\begin{center}
\epsfxsize=14cm
\epsfysize=20cm
\centerline{\epsfbox{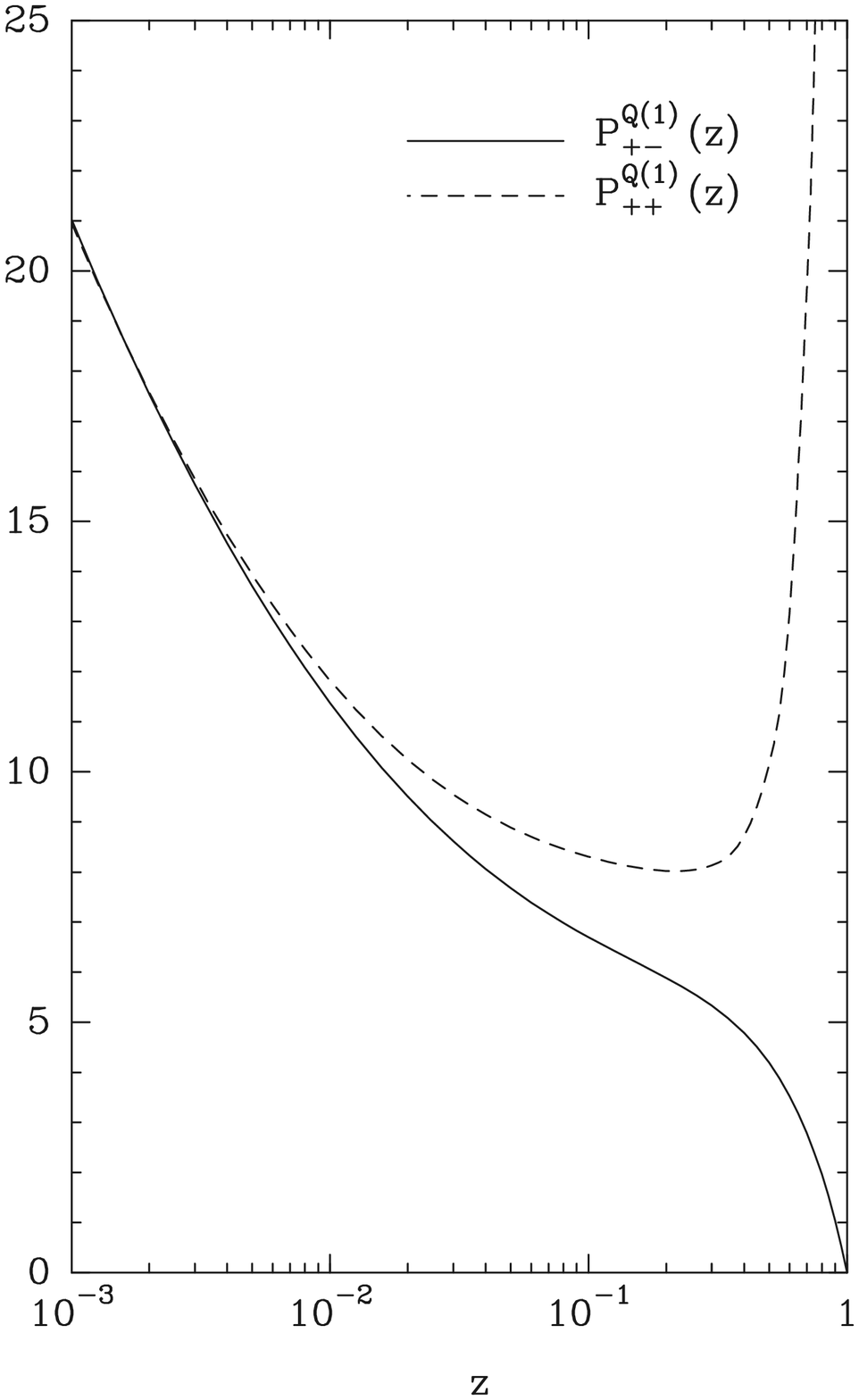}}
\caption{The 'super'-kernels at NLO $P^{Q(1)}_{++}(z)$ and
$P^{Q(1)}_{+-}(z)$ versus $z$.}
\label{fi:dsx}
\end{center}
\end{figure}

Let us make a brief comparison with recent works in this
direction.
The LO analysis of \cite{bar} emphasizes in fact the positivity of
$P^Q_{+-}(z)$ (whose validity is most nontrivial for valence quarks)
and the role played by singlet terms,
while the consideration of $P^Q_{++}(z)$ and, in particular,
of singular terms, are absent.
The same remark applies to  the recent NLO analysis \cite{Vog}
\footnote{The correct NLO $Q^2$ evolution of $h_1$, with no specific
concern to the validity of Soffer inequality, has been also derived in
\cite{Haya}.}
,where however, one can find the general reason for the absence of singular
terms in non-diagonal kernel $P^Q_{+-}(z)$ at NLO,
which is of crucial importance for our proof.
A partial NLO analysis was also performed  in \cite{Cont},
where the specific role of the coefficient function was studied.
The full analysis of the NLO case can now be treated in a very simple way.
We just have to examine at NLO the 'super'-kernels considered above. This requires only the knowledge of $P_{qq}^{(1)}$ and $P_h^{(1)}$ (denoted
$\Delta_TP_{qq}^{(1)}$ in ref.~\cite{Vog}) and according to our
previous analysis, Soffer inequality will be preserved at any
$Q^2 \geq Q^2_{0}$ provided
$P^{Q(1)}_{++}(z)  =  [P_{qq}^{(1)}(z) + P_h^{(1)}(z)]/2$
and $P^{Q(1)}_{+-}(z)  =  [P_{qq}^{(1)}(z) - P_h^{(1)}(z)]/2$ are
positive for $z <1$, while the singular term is occuring only in the
diagonal kernel. All these features appear clearly in the results
of the numerical calculations shown in Fig.~\ref{fi:dsx}, where we
have used the results of \cite{Vog} which correspond to performing
the NLO evolution in the $\overline{{\mbox{\rm MS}}}$ scheme.

One may also consider the analogue of the inequality (\ref{S})
for the complete NLO calculation including the coefficient function
for the Drell-Yan (DY) process \cite{Vog,Cont}.
This corresponds to the choice of the DY factorization scheme, when
the DY cross-section is preserving its Born form \cite{Cont},
while the whole
NLO result is assigned to the evolution of the parton distributions.

 Let us note, however, that Eq.~(\ref{S}) should not be necessarily
valid for the whole NLO result. The reason is that it is not related
to the positivity of the density matrix of a real particle,
but rather to that of a quark in the hadron.
This is in contrast to the well known positivity relation in polarized
Deep Inelastic Scattering (DIS) \cite{Don} $A_2 \leq \sqrt{R}$,
where the observed virtual photon is considered instead.

Although only the full NLO result is physically meaningful,
this does not contradict the fact that there are relations,
like (\ref{S}),
which are valid only partially, and should be
considered like a restriction for the choice of the factorization 
scheme.

For example, all the sum rules coming from the conservation of
some operator (like fermion number or momentum conservation),
generally speaking, are not valid for the full NLO result, and the
factorization scheme should be chosen in order to preserve them.

\begin{figure}
\begin{center}
\epsfxsize=14cm
\epsfysize=20cm
\centerline{\epsfbox{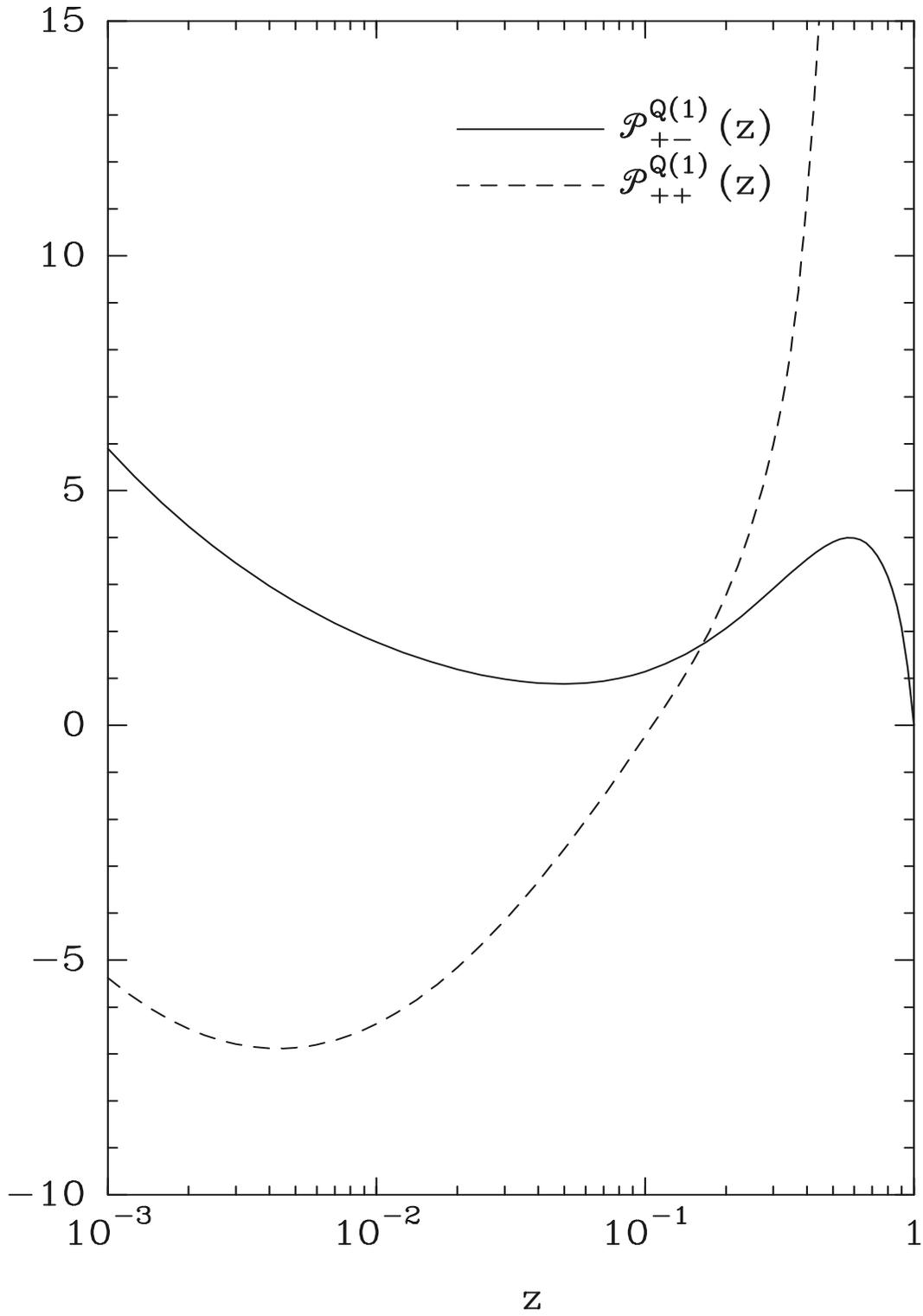}}
\caption{The 'super'-kernels at NLO ${\cal P}^{Q(1)}_{++}(z)$ and
${\cal P}^{Q(1)}_{+-}(z)$ including the DY coefficient functions
versus $z$.}
\label{fi:tdsx}
\end{center}
\end{figure}

\begin{figure}
\begin{center}
\epsfxsize=14cm
\epsfysize=20cm
\centerline{\epsfbox{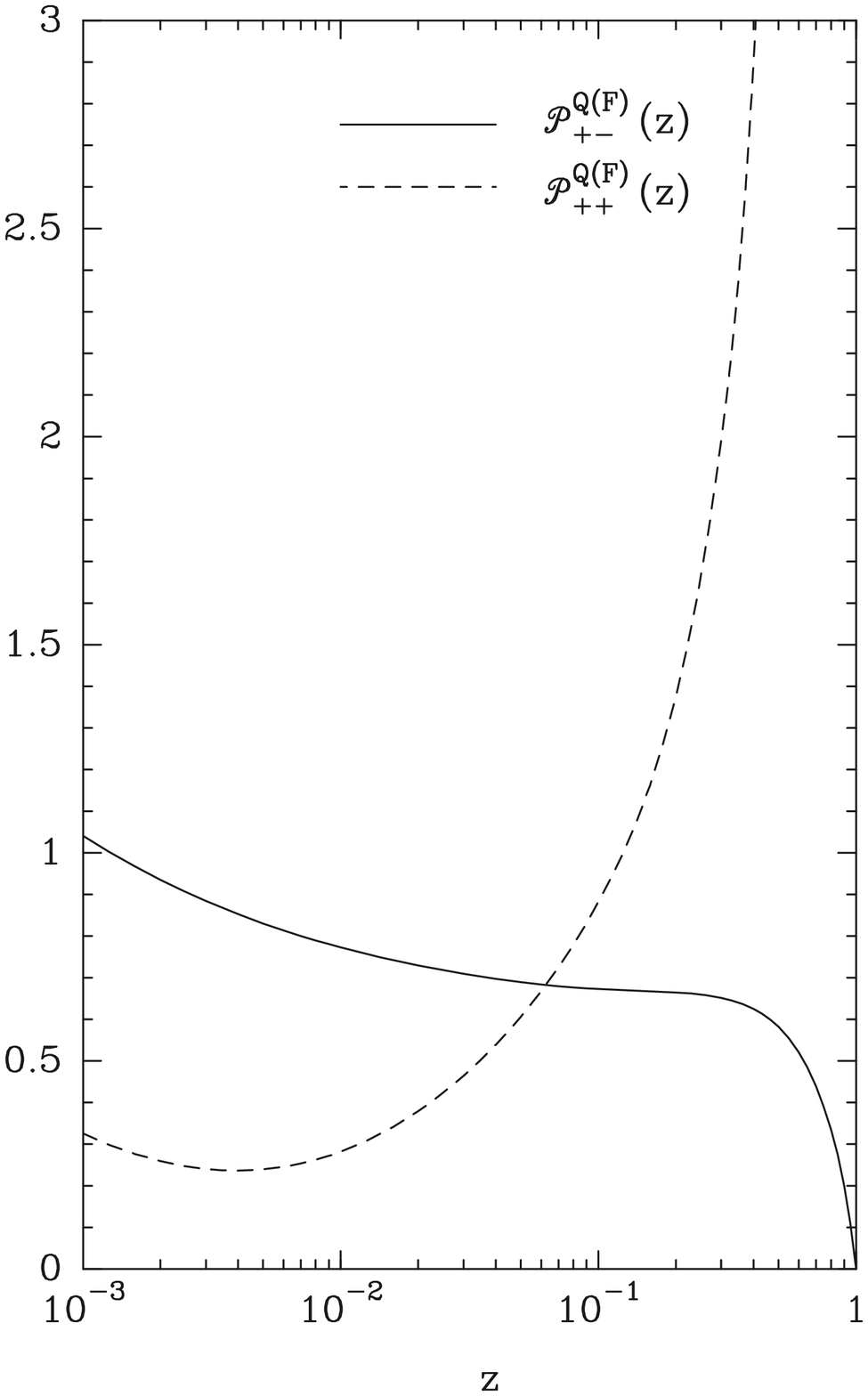}}
\caption{The combined LO and NLO kernels (see Eq.~\ref{pqf})
versus $z$ with $\alpha_S = 0.4$.}
\label{fi:pfdsx}
\end{center}
\end{figure}

We show in Fig.~\ref{fi:tdsx} the calculations of the 'super'-kernels
at NLO including the DY coefficient functions from \cite{Vog}
${\cal P}^{Q(1)}_{++} \hbox{and}~{\cal P}^{Q(1)}_{+-}$.
One observes that while the non-diagonal one remains positive,
this is not the case for the diagonal one ${\cal P}^{Q(1)}_{++}$
which is negative in the small $z$ region. However, if we now
combine LO and NLO, and consider
\begin{eqnarray}
{\cal P}^{Q(F)}_{++}(z) &=& P^{Q}_{++}(z) + \frac{\alpha_S}{2 \pi}
{\cal P}^{Q(1)}_{++}(z), \nonumber \\
{\cal P}^{Q(F)}_{+-}(z) &=& P^{Q}_{+-}(z) + \frac{\alpha_S}{2 \pi}
{\cal P}^{Q(1)}_{+-}(z),
\label{pqf}
\end{eqnarray}
we see in Fig.~\ref{fi:pfdsx} that positivity is preserved even in
the DY factorization scheme.

 In conclusion, we have presented the general proof of positivity
of parton distributions, based on the kinetic interpretation of the
evolution equations.
While the general reasoning is valid at any order of perturbation
theory, the consideration of NLO requires to check the inequalities
for the splitting kernels, for a particular factorization scheme
\cite{BLT,Vog}.
Positivity is, generally speaking, not conserved
for the 'back' evolution to lower $Q^2$, providing a 'time arrow' and
Soffer inequality is proven to be preserved by the LO and NLO evolutions.

\vspace*{7mm}
The authors are deeply indebted to Prof. E. Leader, for drawing their
attention to the problem of positivity as an extra constraint for the 
choice of the factorization scheme in NLO evolution, and also
for helpful discussions, valuable comments and advices.
We would like to acknowledge useful discussions with A.P. Contogouris,
B. Kamal, Z.~Me\-re\-ba\-shvili, S.V. Mikhailov, A. Sch\"afer and
W. Vogelsang.
Two of us,
(C.~B. and O.V.~T.) are  grateful to the Physics Department
of Birkbeck College,
where the essential part of this work was performed,
for his warm hospitality,
and
O.V.~T. thanks the Royal Council in Particle
Physics and Astronomy for financial support.
This research is partly performed in the framework of the Grants
96-02-17631 of Russian Foundation for Fundamental Research
and Grant $N^o_-$ 93-1180 from INTAS.

\newpage

\newpage

\newpage

\begin{thebibliography}{99}

\bibitem{S}
        J. Soffer, Phys. Rev. Lett. {\bf 74} (1995) 1292.

\bibitem{GS} T. Gehrmann and W. Stirling, Phys. Rev. {\bf D53} (1996) 6100.

\bibitem{GRSV} M. Gl\"uck, E. Reya, M. Stratmann and W. Vogelsang,
Phys. Rev. {\bf D53} (1996) 4775.

\bibitem{GL}
V.N. Gribov and L.N. Lipatov, Sov. J. Nucl. Phys. {\bf 15} (1972) 438;
\\ L.N. Lipatov, Sov. J. Nucl. Phys. {\bf 20} (1974) 94;\\
A.P. Bukhvostov, L.N. Lipatov and N.P. Popov, Sov. J. Nucl. Phys.
{\bf 20} (1974) 287.

\bibitem{AP}
G. Altarelli, G. Parisi, Nucl. Phys. {\bf B126} (1977) 298.

\bibitem{D}
Yu.L. Dokshitzer, Sov. Phys. JETP {\bf 46} (1977) 641.


\bibitem{CQ}
        J.C. Collins and J. Qiu,  Phys. Rev. {\bf D39} (1989) 1398.

\bibitem{DP}
        L. Durand and W. Putikka,  Phys. Rev. {\bf D36} (1987) 2840.

\bibitem{OT}
        O.V. Teryaev, in preparation.

\bibitem{BLT}
     C. Bourrely, E. Leader and O.V. Teryaev, in Proceedings
     of VII International Workshop on spin effects in High Energy Physics,
     Dubna, 1997, to appear; abstracts (Dubna Report E2-97-204) -- p. 6.

\bibitem{bar} V. Barone, Univ, Torino Preprint DFRR-68-96,
                        hep-ph/9703343.

\bibitem{Vog} W. Vogelsang, CERN Preprint CERN-TH797-132,
                        hep-ph/9706511.

\bibitem{Haya} S. Kumano and M. Miyama,
              Phys. Rev. {\bf D56} (1997) 2504;
              A. Hayashigaki, Y.~Kawazawa and Y.~Koike,
              hep-ph/9707208 (July 1997).

\bibitem{Cont} B. Kamal, A.P. Contogouris and Z. Merebashvili,
               Phys. Lett. {\bf B376} (1996) 290.

\bibitem{Don} M.G. Doncel and E. de Rafael, Nuovo Cimento {\bf 4A}
              (1971) 363.


\end{thebibliography}
\end{document}